\documentclass[aps, twocolumn,noshowpacs,preprintnumbers,amsmath,amssymb]{revtex4}

\usepackage{verbatim}
\usepackage{wasysym}

\usepackage{graphicx}
\usepackage{dcolumn}
\usepackage{bm}
\usepackage{wrapfig}

\begin{document}

\title{Thermoelectric properties of lead chalcogenide core-shell nanostructures}

\author{Marcus Scheele}
\author{Sven-Ole Peters}
\author{Alexander Littig}
\author{Andreas Kornowski}
\author{Christian Klinke}
\author{Horst Weller}
\affiliation{Institute of Physical Chemistry, University of Hamburg, 20146 Hamburg, Germany}
\author{Niels Oeschler}
\author{Igor Veremchuk}
\affiliation{Max Planck Institute of Chemical Physics of Solids, Noethnitzer Strasse 40, 01187 Dresden, Germany}

\begin{abstract} 

We present the full thermoelectric characterization of nanostructured bulk PbTe and PbTe-PbSe samples fabricated from colloidal core-shell nanoparticles followed by spark plasma sintering. An unusually large thermopower is found in both materials, and the possibility of energy filtering as opposed to grain boundary scattering as an explanation is discussed. A decreased Debye temperature and an increased molar specific heat are in accordance with recent predictions for nanostructured materials. On the basis of these results we propose suitable core-shell material combinations for future thermoelectric materials of large electric conductivities in combination with an increased thermopower by energy filtering. 

\end{abstract}

\maketitle

Lead chalcogenides have been extensively investigated for thermoelectric applications \cite{1}. With a maximum of 0.7 (at 773 K) in the thermoelectric figure of
merit (ZT), PbTe is one of the best thermoelectric materials at intermediate temperatures (at 450-800 K) \cite{2}. To increase ZT toward economical competitiveness with conventional electric power generators, nanostructuring is nowadays widely applied \cite{3}. A key concept of this approach is the phononglass electron-crystal (PGEC) effect, which arises from the preferential scattering of phonons over charge carriers leading to a larger decrease in thermal conductivity ($\kappa$) than in electric conductivity ($\sigma$) \cite{4}. Since ZT is proportional to $\sigma$/$\kappa$, the PGEC effect leads to an enhancement in ZT \cite{5}. Nanoparticle inclusions of a second material (e.g., Sb$_2$Te$_3$, AgSbTe$_2$, or Sb) into PbTe are known to lead to drastic reductions in $\kappa$ \cite{6,7,8}. Specifically, inclusions of PbS have been shown to maintain high electron mobilities due to the presence of coherent interfaces but decrease phononic contributions to $\kappa$ to 0.5 Wm$^{-1}$K$^{-1}$ for temperatures of 300-700 K \cite{9,10}. This is believed to be the lower limit for PbTe-based materials \cite{11}. Therefore, complementary
approaches to the PGEC-concept need to be addressed to achieve further enhancements in ZT.

A promising attempt is energy filtering by intentionally introducing potential barriers to charge carrier transport. Since most materials applied for thermoelectrics are degenerate semiconductors, the Fermi level (E$_F$) in these compounds is located close to or even inside a band. This leads to the disadvantageous situation that carriers from both sides of the Fermi level contribute to the total thermopower. Since thermopower is ameasure for the average energyper charge carrier with respect to E$_F$, the individual thermopower of carriers from one side of E$_F$ partially cancel that of carriers from the other side of E$_F$. Thus, the total thermopower
of a degenerate semiconductor would be larger if carriers from one side of E$_F$ are immobilized. When a low potential barrier ($\Delta$E $<$ 100 meV) is introduced above E$_F$, only carriers of higher energy can contribute to electric transport and the average energy per charge carrier is increased. This concept has been modeled for metal-based superlattices and nanocomposites \cite{12,13}. Experimental proof of principle has been provided by samples of nanostructured bulk PbTe. The large surface area in such nanograined materials provides a high trap state density in each grain caused by surface adsorbates. When charge carriers are trapped on the surface, they provide an energy barrier to electric transport which leads to energy filtering \cite{14}. However, it is unclear if the observed increase in thermopower in nanograined materials is due to energy filtering or to an increase in the scattering parameter caused by the large density of incoherent crystalline domains \cite{15}.

Here, we study the effect of epitaxially growing a shell of a secondmaterial onto PbTe nanoparticles and fabricate macroscopic nanostructured samples of this material. We show that core-shell nanoparticles are ideal candidates to study the energy filtering concept and increase a material's thermopower. Alloying of the core and shell material allows the formerly heterophased grains to transform into a single phase of similar grain size, which allows us to determine whether the increased thermopower is a result of the core-shell structure or the limited grain size.

\begin{figure*}[htbp]
  \centering
  \includegraphics[width=0.98\textwidth]{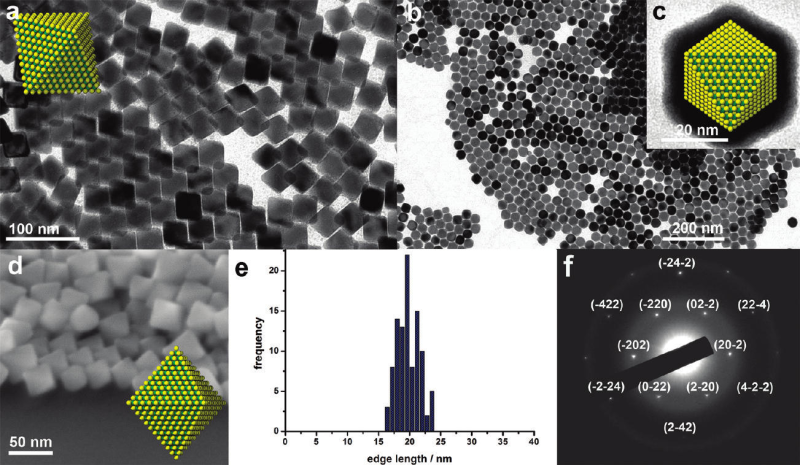}
  \caption{\textit{(a-c) TEM images of octahedral PbTe nanoparticles in (a) $<$110$>$- and (b and c) $<$111$>$-projection. (d) SEM image of PbTe nanoparticles. (e) Histogram of the PbTe nanoparticles' edge lengths. (f) SAED of the single PbTe nanoparticle in panel c. Indexing is according to the rock salt structure.}}
\end{figure*}

\subsection*{RESULTS AND DISCUSSION}

PbTe nanoparticles were synthesized in diphenylether solution on applying oleic acid as the stabilizing ligand. A typical synthesis yields several 100 mg per batch of monodisperse octahedral nanoparticles (Figure 1a-f). A histogram of the edge lengths for the particles displayed in Figure 1b is given in Figure 1e, revealing an average of 20 nm. If assuming a regular octahedra, this corresponds to an insphere diameter of 16 nm.

In Figure 1f, the electron diffraction pattern of the single PbTe crystal displayed in Figure 1c is depicted revealing the direction of view as $<$111$>$. Since PbTe
crystallizes in the fcc lattice, it follows that each facet of the octahedron is a $\{$111$\}$-facet. Within the rocksalt structure, structural discrimination from cubic to octahedral can be achieved by thermodynamic versus kinetic control, respectively. Slow growth kinetics of the $\{$111$\}$-facets are typically achieved for a large excess of Pb-precursor in combination with the presence of primary amines or thiols \cite{16,17}. As a reason for this unusual stability of $\{$111$\}$-facets, a $\mu$-Pb3-SR-bonding at least in the case of thiol-stabilization was proposed. $\{$111$\}$-Facets in PbTe are composed exclusively of either Pb or Te atoms which distinguishes them from $\{$100$\}$-facets in which both atoms are equally abundant. Only in the former case, multiple binding modes with high binding energies are possible. When working with a large excess of Pb-precursor, the $\{$111$\}$-facets are likely to be terminated by Pb-atoms in which case
the growth in this direction can be inhibited.

Another important aspect about structural discrimination in lead chalcogenide nanocrystals has been investigated by Houtepen \textit{et al.}: the presence of catalytic amounts of acetic acid \cite{18}. We can fully confirm their observations insofar as the presence of a small amount of acetic acid is essential in order to obtain the octahedra depicted in Figure 1a-d. If acetic acid is absent, only cubic nanoparticals can be synthesized under otherwise identical conditions.

To allow for high electric conductivities, all organic residues have to be removed from the inorganic nanoparticles. Oleic acid stabilized PbTe nanoparticles
in hexane were treated with an excess of phosphonic acid in solution to induce ligand exchange. Phosphonic acid ligands were removed by adding a methanolic
ammonia solution which separated the polar nanomaterial from the nonpolar organic residues following a previously developed protocol \cite{19}. After drying the inorganic material under vacuum, a dark-gray nanopowder was obtained.

\begin{figure*}[htbp]
  \centering
  \includegraphics[width=0.98\textwidth]{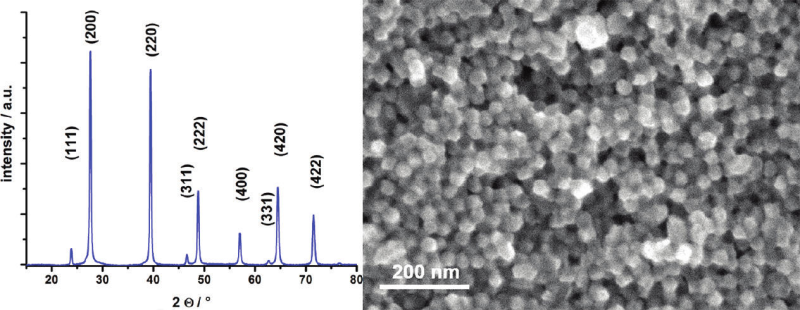}
  \caption{\textit{(left) XRPD pattern of spark plasma sintered pellet of PbTe nanoparticles. Indexed according to the rock salt structure. (Right) SEM image of the compacted pellet of PbTe nanoparticles.}}
\end{figure*}

To fabricate a nanostructured bulk material, this powder was spark plasma sintered (SPS) to a macroscopic pellet. Pellets of PbTe nanoparticles were silver-metallic in appearance with a density of 7.30 $\pm$ 0.10 g cm$^{-3}$ (89\% of theoretical density) under the conditions specified in the experimental section. To achieve larger densities toward 100\% of the theoretical value, significantly higher sintering temperatures and duration are required which resulted in undesirable grain
growth during compaction.

\begin{figure*}[htbp]
  \centering
  \includegraphics[width=0.98\textwidth]{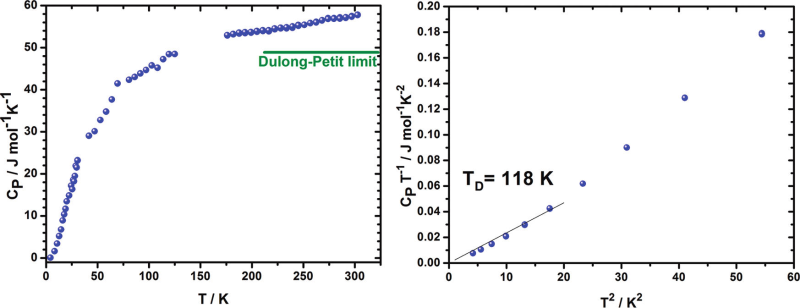}
  \caption{\textit{(left) Specific heat capacity of spark plasma sintered PbTe nanoparticles. (Right) Plot of C$_P$/T vs T$^2$ to derive the Debye temperature of a pellet of spark plasma sintered PbTe nanoparticles.}}
\end{figure*}

We note that several other groups have compacted lead telluride nanocrystals before by either SPS or hotpressing. In all cases, the applied temperatures were
significantly higher (310-500$^{\circ}$C) than our applied temperature range (100-200$^{\circ}$C) \cite{5,8,15,20}. Our results suggest that nanostructured PbTe bulk samples with grain sizes $<$30 nm (Figure 2) can only be obtained if the temperature during SPS is kept close to 100$^{\circ}$C. At 195$^{\circ}$C and even at 154$^{\circ}$C significant grain growth was observed. (Note that during SPS the actual sample temperature inside the die is not known but anticipated to be somewhat higher.)

The low melting point of nanostructured PbTe pellets with grain sizes of $\sim$30 nm is a consequence of the high surface to volume ratio in this material. Surface
atoms possess higher degrees of freedom and greater energy as opposed to their bulk equivalents due to unsaturated binding sites. This is known to cause a large depression in melting temperature for sufficiently small crystal grains \cite{21}. Note that bulk PbTe melts at 924$^{\circ}$C.

Another effect of the high surface to volume ratio in nanostructured materials is an alteration of the molar specific heat (C$_P$) as visible in Figure 3. There are a
number of experimental verifications that nanograined materials generally possess a specific heat which exceeds Dulong-Petit's rule by up to 40\% at 300 K \cite{22,23,24,25}. Other groups found reduced Debye temperatures (T$_D$) in nanostructured materials \cite{26,27}. In the nanostructured PbTe sample in this work, both effects are present. At 300 K, Dulong-Petit's rule with C$_P$ approaching 3 N R is violated by an excess of almost 20\% (Figure 3a).

From the plot in Figure 3b, TD can be evaluated applying

\begin{equation}
	C_{P} = \frac{12 \pi ^4}{5} N R \left( \frac{T}{T_D} \right)^3
\end{equation}

with N = 2 being the number of atoms per molecule and R = 8.314 J mol$^{-1}$K$^{-1}$ being the gas constant. The resulting T$_D$ = 118 K is significantly smaller than
the value reported for large PbTe single crystals (T$_D$ = 168 K) \cite{28}. These findings qualitatively support a recent theoretic description of the dependence of C$_{P}$ and T$_D$ on the grain size in nanostructured materials \cite{29}.

\begin{figure}[htbp]
  \centering
  \includegraphics[width=0.48\textwidth]{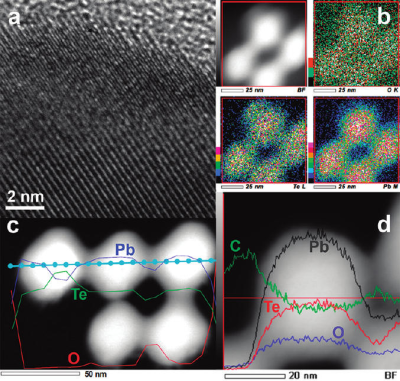}
  \caption{\textit{(a) HR-TEM image of the edge of a single PbTe nanoparticle. The region of lower contrast shows the amorphous carbon substrate. (b) EDXS-based elemental mapping of four PbTe nanoparticles: (top left) Bright field STEM image, (top right) oxygen K-line, (bottom left) tellurium L-line and (bottom right) lead M-line. Scale bars are 25 nm. (c) EDXS line scan across three PbTe nanoparticles with 5 nm resolution. The spatial variation of the quantified EDXS signals of Pb, O, and Te are depicted over the corresponding STEM image in the background. The scale bar corresponds to 50 nm. (d) Continuous, nonquantified EDXS line scan over a single PbTe nanoparticle and spatial variation of the Pb, Te, C, and O signal plotted over the STEM image in the background. The scale bar corresponds to 20 nm.}}
\end{figure}

In Figure 4, the elemental composition of individual PbTe nanoparticles after ligand removal is investigated by high angle annular dark field (HAADF) and energy
dispersive X-ray spectroscopy (EDXS). Even close to the surface, particles appear to be single-crystalline (Figure 4a). Elemental mapping (Figure 4b) and spatially
resolved EDXS scans across individual particles (Figure 4c) show significant amounts of oxygen. Oxidation of PbTe is a complex, multistaged process which is believed to start with the formation of peroxide-like structures and terminate in the formation of PbTeO$_3$ on exposure to large O$_2$ concentrations \cite{30,31}. The absence of features originating from such materials in XRPD (Figure 2 left) and high resolution transmission electron microscopy (HRTEM, Figure 4a) let us speculate
that oxidation by O$_2$ affects predominantly the surface of PbTe.

The continuous nonquantified EDXS scan in Figure 4d reveals the carbon signal originating from the free (carbon-covered) TEM substrate to be larger than that from an area additionally covered with a PbTe nanoparticle. This underlines the effectiveness of the ligand removal process. 

The thermoelectric properties of sintered pellets of PbTe nanoparticles as imaged by scanning electron microscopy (SEM, Figure 2 right) are displayed in Figure 5. In accordance with the picture of surface-oxidized PbTe nanoparticles, electric transport (Figure 5a) shows strong temperature activated behavior and a resistivity ($\rho$) which is more than 1 order of magnitude larger than in a comparable bulk material \cite{10}. Nolas and co-workers found a similar behavior for nanostructured PbTe samples and explained it with grain boundary potential barrier scattering due to chemisorbed oxygen at the PbTe grain boundaries \cite{13,14}. They calculated and experimentally demonstrated that surface oxygen can trap charge carriers and provide an energy filter which allows only carriers of sufficient energy to pass and contribute to electric conduction as well as thermopower (S). Since this increases the mean energy per carrier, they found significantly larger S values. Such energy filtering is especially effective in heavily doped semiconductors where the Fermi level (E$_F$) is often found almost inside a band. At temperatures sufficiently above absolute zero, a substantial amount of carriers with energies greater and smaller than E$_F$ counteract each other leading to a partial cancellation in S. Trapping low-energy carriers at grain boundaries prevents such cancellations and increases the total thermopower \cite{32}. The large thermopower of nanostructured surface oxidized PbTe with grain sizes of $\sim$30 nm presented in Figure 5b may be a signature of this effect. With 625 $\mu$V K$^{-1}$ at room temperature, this significantly exceeds the value of 265 $\mu$V K$^{-1}$ of typical bulk samples and that of the Nolas group with 325 $\mu$V K$^{-1}$ for grain sizes of 350 nm \cite{14}. Further, Heremans \textit{et al.} reported nanostructured PbTe samples of 40 nm grain size with 500 $\mu$V K$^{-1}$ in thermopower \cite{15}. Although a complete comparison of thermopower values of these samples also requires carrier concentrations, one can see a general trend to larger thermopower as the crystal size decreases. This could be understood in terms of the increasing surfaces-to-volume ratio and thus the larger number of trap states with smaller crystalline domains. In the present case, a larger trap state density due to surface oxidation raises the barrier and hole transport from grain to
grain is greatly impeded by the frequent occurrence of narrow sinks in the valence band edges of each grain. This reduces the amount of mobile holes capable of traveling through the entire nanostructured sample. Meanwhile the average energy per hole increases as only holes of sufficient energy can pass the barriers. As
a consequence $\mu$ and S would be expected to show the increase discussed above.

\begin{figure*}[htbp]
  \centering
  \includegraphics[width=0.98\textwidth]{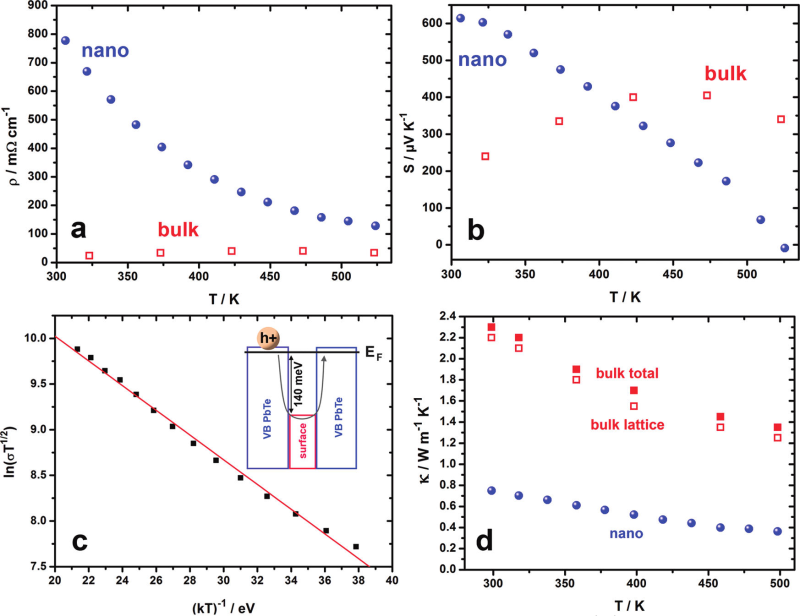}
  \caption{\textit{(a) Resistivity, (b) thermopower, and (d) thermal conductivity of spark plasma sintered, surface-oxidized PbTe nanoparticles (circles) in comparison to bulk values reported in literature (squares). The bulk thermal conductivity in (d) is plotted in terms of its lattice contribution (open squares) and total value (closed squares). The lattice contribution of the nanostructured sample is identical with the total thermal conductivity. (c) Linear fit to the plot of ln($\sigma$ T$^{-1/2}$) vs. 1/(kT) to derive the energy barrier associated with grain boundary potential barrier scattering. The inset sketches the electronic structure at the valence band edges between two grains.}}
\end{figure*}

We note that the large thermopower reported in this work could also be due to a very low concentration of free carriers in the PbTe species. Following Crocker \textit{et al.} this would be the case for n $\leq$ 1 x 10$^{17}$ cm$^{-3}$ \cite{33}. However, it is hard to imagine how such a material could show the breakdown in thermopower as displayed in Figure 5b. In fact for pure PbTe, the bandgap even widens in this temperature regime so that one usually finds an increase in S until T $\approx$ 450 K where a hole pocket at the $\Sigma$ point becomes the principal valence band maximum. Thus, we argue that an additional mechanism must be operative to account for the temperature dependence of S reported in this work. We believe that energy filtering by a small potential barrier is a likely explanation.

To derive the height of the energy barrier at the grain boundaries (E$_B$), we follow J. Y. W. Seto and assume an effective electric conductivity according to the following equation \cite{34}:

\begin{equation}
	\sigma _{\text{eff}} \propto T^{-1/2} \exp \left[ - \frac{E_{B}}{k T} \right]  
\end{equation}

When plotting ln($\sigma$ T$^{-1/2}$) vs. 1/(kT) in Figure 5c, the straight line indicates that this assumption is justified and yields a height for E$_{B}$ of 140 meV. This should be compared to the material used by the Nolas group with E$_{B}$ = 60 meV.

Figure 5d verifies the well-known advantages of nanostructured materials in terms of minimizing thermal conductivity ($\kappa$). Owing to poor electric transport
properties, the depicted total thermal conductivity of the nanostructured material is practically identical to the lattice contribution. Note that both, resistivity (Figure 5a) and thermal conductivity, have been corrected for effects of porosity according to the Maxwell-Eucken approach (parameters: porosity (P) = 11\% and $\beta$ = 2) as described elsewhere \cite{35}. In comparison with bulk PbTe, the lattice contribution to the thermal conductivity is reduced by approximately 70\%
over the whole temperature range between 300 and 500 K. The lowest reported lattice thermal conductivity of PbTe was found in a highly insulating nanostructured
pellet with grains of 9-12 nm with 0.5 W m$^{-1}$K$^{-1}$ at 300 K (not corrected for porosity) \cite{11}. This is only slightly smaller than the value reported in this work (0.75 W m$^{-1}$ K$^{-1}$) demonstrating that below a certain grain size only small additional reductions in $\kappa$ can be achieved.

\begin{figure}[htbp]
  \centering
  \includegraphics[width=0.48\textwidth]{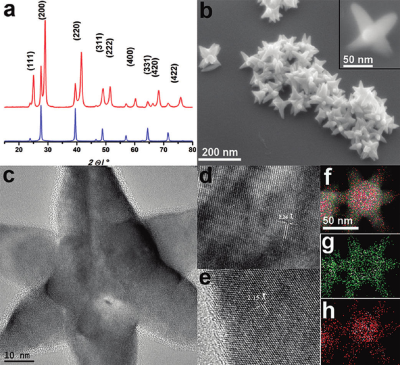}
  \caption{\textit{(a) XRPD of PbTe-PbSe nanostars (top) and PbTe nanoparticles (bottom). Indexed according to the rock salt structure. (b) SEM images of PbTe-PbSe nanostars. (c-e) TEM images of a single nanostar, its center and edge. (f-h) EDXS-based elemental mapping of an individual nanostar. From top to bottom: Bright field STEM image with elemental mapping overlay, tellurium L-line, and selenium K-line.}}
\end{figure}

To conclude this discussion, we note that the PbTe nanocrystals applied in this work are too small to result in large ZT values. The trap state densities due to
chemisorbed oxygen result in a large increase in resistivity which is not fully compensated by the increase in thermopower. To optimize the interplay of these two parameters, lower barriers should be applied. This could be achieved by increasing the crystal size or growing a shell of a second material with suitable band offset onto the oxygen-free surface of the PbTe nanocrystals. 

Following this idea, we have synthesized a second material: PbTe-PbSe core-shell nanostars as presented in Figure 6. PbSe has the same crystal structure as PbTe and almost the same band gap. Since a useful barrier height for effective energy filtering in combination with large electric conductivity is believed to be 40-100 meV, combining two materials of very similar band gaps to an alternating heterostructure can be highly advantageous for thermoelectric applications due to an expected increase in the power factor (S$^2$$\sigma$) \cite{12}. XRPD analysis (Figure 6a) reveals the formation of a heterophased crystalline material which could be indexed to cubic PbTe and PbSe (Fm-3m). Each index refers to a pair of reflections as both crystalline phases are of the same space group and possess the same number of reflections at similar angular positions. For the higher indexed planes, some reflections of the two phases overlap and appear unresolved. For comparison, the XRPD pattern of the starting PbTe nanoparticles is also displayed. 

SEM imaging (Figure 6b) shows the three-dimensional six-armed nanostar structure. This can be understood in terms of epitaxial growth of PbSe in the $<$100$>$-direction onto each of the octahedron's six tips. We believe the key to this anisotropic growth to be the strong dipole in this direction caused by the $\{$111$\}$-facets. Each star measures roughly 75 nm from tip to tip.

HRTEM images of a single nanostar (Figure 6c) display single-crystallinity and a difference in lattice spacing (d) as one compares areas close to the core (Figure 6d) with areas near the edge of an arm (Figure 6e). The predominant spacing in the core (d = 2.26 $\pm$ 0.03 \AA) is substantially larger than the spacing found in the arms (d = 2.15 $\pm$ 0.03 \AA). This should be compared to the lattice spacing of the (220)-planes in PbTe (2.28 \AA) and PbSe (2.16 \AA) reported in literature \cite{36}. We could not locate an area with a significant number of stacking faults characteristic for an abrupt change from pure PbTe to pure PbSe crystallinematerial. We assume an intermediate section where a PbTe1-xSex alloy is present.

\begin{figure*}[htbp]
  \centering
  \includegraphics[width=0.98\textwidth]{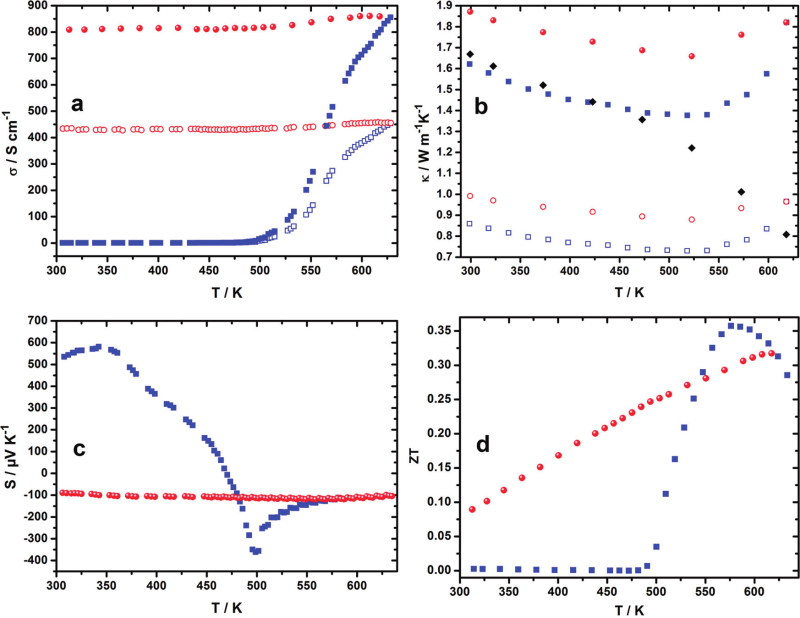}
  \caption{\textit{(a) Electric conductivity, (b) thermal conductivity, (c) thermopower, and (d) thermoelectric figure of merit of spark plasma sintered pellets of surface-oxidized PbTe-PbSe nanostars. All measurements have been carried out in two consecutive cycles. After the first measurement (squares), the pellets were cooled to room temperature and the measurement was repeated (circles). The raw measurements of $\sigma$ and $\kappa$ (open symbols) have been corrected for porosity (full symbols) as explained in the text. The lattice contribution to $\kappa$ (diamonds) has been calculated from the second measurement as explained in the text.}}
\end{figure*}

Figure 6 panels f-h display elemental mapping of an individual nanostar by EDXS with respect to tellurium and selenium. Tellurium is almost exclusively found in the core, whereas selenium is enriched in the arms. Note that due to the three-dimensional structure there will always be one arm located behind the core of the nanostar which misleadingly adds to the selenium signal recorded in the core area.

Comparable to what we have found for pure PbTe nanoparticles, PbTe-PbSe core-shell nanostars appear to be oxidized, too. It is difficult to accurately access the degree of oxidization, but the absence of oxide reflections in XRPD analysis let us speculate that it is mainly the surface which is affected. This is supported my EDXS line scans (not shown here) similar to what has been displayed for pure PbTe in Figure 4c.

PbTe-PbSe nanostars were treated according to the procedure described for PbTe nanooctahedra to remove organic ligands and prepare a nanopowder for SPS. The density after SPS compaction was 6.40 $\pm$ 0.10 g cm$^{-3}$ (78\% of theoretical density). This relatively low value accounts for the bulky starlike structure which was preserved after SPS. Nonetheless, samples were mechanically stable and silver metallic in appearance.

The thermoelectric properties of sintered pellets of PbTe-PbSe nanostars are presented in Figure 7. Electric ($\sigma$) and thermal conductivity ($\kappa$) have been corrected for porosity as described for pure PbTe (P = 22\%, $\beta$ = 2). Owing to the large degree of porosity, these corrections were so significant that we prefer to show them together with the actually measured values for comparison. Thermopower (S) is known to be practically unaffected by porosity and was not corrected \cite{37}. On using the measurements of $\sigma$, $\kappa$, and S, ZT was calculated according to

\begin{equation}
	ZT = \frac{\sigma S^2}{\kappa} T
\end{equation}

Since the corrections for porosity in $\sigma$ and $\kappa$ cancel in ZT, a correction to ZT is not necessary. Each measurement was performed in three cycles over the temperature range from 300 to 625 K. The first two cycles are shown for each thermoelectric parameter. The third cycle was practically identical to the second cycle in all cases and is not shown.

Similar to the pellets of sintered PbTe nanoparticles (Figure 5), the pellets of sintered PbTe-PbSe nanostars show an equally low electric conductivity. This is most
likely due to surface oxidation, too. Any energy filtering effect from a small band edge offset between PbTe and PbSe would be invisible in Figure 7a since the surface oxide barrier is by far the highest barrier.

\begin{figure}[htbp]
  \centering
  \includegraphics[width=0.48\textwidth]{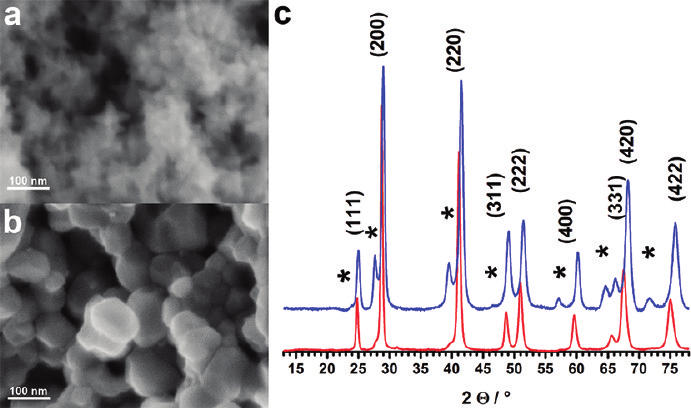}
  \caption{\textit{SEM images of spark plasma sintered pellets of surface-oxidized PbTe-PbSe before (a) and after (b) the first measurement cycle. (c) XRPD of the pellet before (top) and after (bottom) the first measurement cycle. Indexed according to the rock salt structure. Reflections due to a pure PbTe phase are marked with an asterisk.}}
\end{figure}

Although $\kappa$ (Figure 7b) appears to be low, the reduction is predominantly due to porosity which becomes apparent after applying a correction for porosity. With
1.6 W m$^{-1}$ K$^{-1}$, thermal conductivity at room temperature is twice as large as in the case of the nanostructured PbTe samples in Figure 5d, which may be attributed to the larger particle diameter (75 nm vs. 30 nm). The remaining reduction in $\kappa$ compared to bulk PbTe (2.2 W m$^{-1}$ K$^{-1}$) is due to combined effects of crystal boundary and ion impurity scattering.

Thermopower measurements (Figure 7c) reveal a remarkable temperature dependence of the as-sintered PbTe-PbSe nanostars. At room temperature, S is practically identical to that of nanostructured surface oxidized PbTe (Figure 5b). Presuming a surface oxide barrier of equal height, this result had to be expected. However, S decreases even more rapidly than for nanostructured PbTe, changes it sign to n-type at 470 K, reaches a sharp turning point at 500 K with a maximum of -360 $\mu$V K$^{-1}$ and decreases to -100 $\mu$V K$^{-1}$ above 575 K. This should be viewed in the light of a sharp improvement in electric conductivity (Figure 7a) beginning at 500 K which terminates in an increase in $\sigma$ by almost 2 orders of magnitude. We explain this with two simultaneously operative processes:

\begin{enumerate}
\item 

Thermally induced alloying of the two chalcogenide phases yields a single alloyed phase with degenerate semiconductor characteristics. Note that selenium is a good n-type dopant for PbTe. As the temperature rises, an increasing number of donor states are created close to the conduction band edge which is consecutively populated with carriers. This leads to a decrease in S since holes and electrons counteract each other and cancel their contribution to thermopower. At 470 K the total S is zero but $\sigma$ is still low since the amount of conduction band electrons needed to compensate those holes with sufficient energy to pass the oxide barrier between the valence bands is small. Between 470 and 500 K, consecutive alloying lifts E$_F$ close to the conduction band edge where the density of states is now quite large. This leads to a maximum in S and a beginning increase in $\sigma$. However, this decreases S since a significant number of mobile electrons on both sides of E$_F$ is now available which cancel each other in their contribution to the total thermopower.

\item 

Thermal annealing destroys the thin oxide layer on the surfaces of the material's grains forming mere oxide islands within the conductive PbTe$_{1-x}$Se$_x$ matrix. This improves $\sigma$ and decreases S since the potential barriers vanish and no more energy filtering can take place. For the changes in microstructure due to
thermal annealing see Figure 8.

\end{enumerate}

We note a recent work by Zhang \textit{et al.} reporting on n-type doping of formerly p-type PbTe nanocrystals with Bi$^{3+}$ as the dopant \cite{38}.

To test the material's changed transport characteristics upon heating we have performed a second measurement over the entire temperature range after cooling down to room temperature. S and $\sigma$ were almost independent of T and were reproducible on performing a third test run to verify performance stability in this temperature window. It can be concluded that the large temperature dependence of S and $sigma$ in the first measurement cycle ismainly ameasure for the thermally induced
solid-state reaction which vanishes once the solid solution has been formed. Note that upon correcting $\sigma$ for porosity, a room temperature value of over 800 S cm$^{-1}$ is obtained which may be compared to 1200 S cm$^{-1}$ of a bulk solid solution of PbTe$_{0.7}$S$_{0.3}$ published recently \cite{10}.

$\kappa$ is less affected by the alloying process and merely displays a moderate increase above 525 K due to the increasing contribution by electron transport ($\kappa_e$). What appears to be a gain in $\kappa$ when comparing the first and second measurement cycle is mostly the effect of improved electric conductivity. This becomes apparent as we estimate the lattice contribution ($\kappa_L$) after alloying by applying the Wiedemann-Franz law:

\begin{equation}
	 \kappa_L = \kappa - \kappa_e = L \sigma T
\end{equation}

with L = 2.0 x 10$^{-8}$ V$^2$ K$^{-2}$ being the Lorentz constant. We note that particularly nanostructured materials can show large deviations of L from the value used here for heavily degenerate bulk semiconductors so this estimate should be handled with care \cite{39}. However, it is intriguing to see that $\kappa_L$ thus calculated from the data of the second measurement cycle matches the total $\kappa$ of the first measurement cycle until the point where alloying adds a significant $\kappa_e$ to the otherwise purely phononic total $\kappa$.

We conclude that the thermally induced alloying of PbTe-PbSe nanostars has a very small effect, if any, on $\kappa_L$. This is actually surprising as one would expect a large degree of impurity scattering and thus a decrease in $\kappa_L$.

For completeness we display the thermoelectric figure of merit (ZT, Figure 7d). Once electric conduction improves, ZT rises to up to 0.35 at 570 K which is comparable to bulk PbTe$_{1-x}$Se$_x$. With a repeat of the measurement after alloying, ZT scales monotonously with T to ZT = 0.3 at 625 K.

In Figure 8, we are investigating crystallographic and structural changes in pellets of sintered surface oxidized PbTe-PbSe nanostars under the conditions applied during transport measurements. Figure 8a is an SEM image of the sample before the measurement. Individual nanostars appear to be well preserved and form a porous network. The SEM image in Figure 8b displays the fine structure of the sample after the first measurement cycle. Polyhedral particles of similar size have replaced the former nanostars. It is important to note that the large thermopower of as-sintered surface oxidized PbTe-PbSe nanostars (Figure 7c) is obviously not due to the small grain size since grain boundary scattering by 50-100 nm grains of PbTe$_{1-x}$Se$_x$ is not sufficient to significantly increase S. What does increase S however is a barrier for charge carrier transport which could be due to two distinct effects: (I) an increase in the density of states by quantum confinement or (II) a depletion of the contribution of low-energy charge carriers to total thermopower. Both effects are known to increase S \cite{34,40}. The exciton-Bohr-radii of PbTe
and PbSe are 152 and 46 nm, respectively \cite{41,42}. Thus, in the case of surface-oxidized PbTe in this work, quantum confinement effects may be present, whereas for surface oxidized PbTe-PbSe nanostars this is less probable. Since both materials show roughly the same thermopower, we speculate that quantum confinement does
not play a significant role in either of the two materials. Instead, we believe that energy filtering by immobilizing low-energy holes due to a sink in the valence band edge is the operative mechanism.

It should be noted that most studies about thermopower enhancement by some scattering mechanism at grain boundaries have been carried out with nanostructured lead chalcogenides. Since the effects on S of grain boundary and potential barrier scattering are practically undistinguishable, the correct interpretation is somewhat academic in materials where both features are present. Owing to the easily oxidized surfaces of lead chalcogenide nanostructures, this is the case for most materials studied so far. When extending this principle to other systems however, the findings in the present work have practical implications. We suggest to intentionally introduce suitable energy barriers for charge carrier transport rather than to merely rely on the effect of grain boundary scattering. The latter may be too weak to force significant enhancements in S.

The XRPD patterns in Figure 8c investigate crystallographic changes during this process. For comparison, the XRPD pattern of thermally untreated surface oxidized PbTe-PbSe nanostars from Figure 6a is reproduced in the upper part. Indexing refers to twin reflections of the same lattice plane in PbTe and PbSe, respectively. PbTe reflections of the same index as their PbSe counterparts occur at smaller angles and are marked with an asterisk. In the lower part of Figure 8c, the XRPD pattern of the same sample after transport measurements is displayed. Each pair of reflections has fused into one single reflection of similar intensity positioned at an intermediate angle. This supports our interpretation of transport measurements in terms of a thermally induced formation of a PbTe$_{1-x}$Se$_x$ solid solution. Upon more measurement cycles no significant changes in terms of structure and crystalline composition could be detected.

On the basis of the results of this work, we propose the following core-shell material for future investigations: PbTe$_{1-x}$Se$_x$ with suitable n-type doping as the corematerial and pure PbSe as the shell. If we assume the position of the conduction band edge of the core to be between that of pure PbTe and PbSe, there would be a potential barrier to electron transport between 0 and 100 meV as one moves from the core to the shell based on the calculations of Wei and Zunger \cite{43}. Any exposure to oxygen must be strictly avoided.

\subsection*{CONCLUSION}

We have studied the thermoelectric properties of surface-oxidized, colloidal PbTe nanoparticles by spark plasma sintering them to a nanostructured bulk sample. The oxidized surfaces of each grain provided an effective barrier to hole transport with a height of 140 meV. An unusually large thermopower in combination with low
electric conductivity has been found. The effect of energy filtering as a possible explanation has been discussed.

A similar behavior was found for surface oxidized PbTe-PbSe core-shell nanostars. By thermally induced alloying, this material could be transformed from p-type to n-type on which electric transport improved greatly. The results of this work imply that effective thermopower enhancement in nanostructured bulk materials is driven predominantly by suitable energy barriers and not by the limited grain sizes alone.

\subsection*{ACKNOWLEDGEMENTS}

We thank A. J. Minnich for fruitful discussions. A Ph.D. grant by the \textit{Studienstiftung des deutschen Volkes} is gratefully acknowledged.

\subsection*{EXPERIMENTAL DETAILS}

All manipulations were carried out under an inert atmosphere using standard Schlenck techniques if not stated otherwise.

\textbf{Preparation of a 0.500 M Solution of Tellurium in TOP (Te$@$TOP)(step 1, product (I)).} In a glovebox, tellurium (1.276 g, 10.00 mmol, 99.999\%, Chempur) and tetradecylphosphonic acid (102 mg, Alfa Aesar) were suspended in distilled TOP (20.0 mL, 90\%, Merck) under stirring. The mixture was heated stepwise to 230$^\circ$C from room temperature by increasing the temperature by approximately 50$^\circ$C every 30 min. The final temperature was kept until a completely transparent, orange solution was obtained which turned to bright-yellow on cooling to room temperature. The solution was stored in the glovebox.

\textbf{Synthesis of Octahedral PbTe Nanoparticles (step 2, product (II)).} In a typical synthesis, lead acetate trihydrate (0.785 g, 2.07 mmol, 99\% Aldrich) was mixed with oleic acid (1.50 mL, 70\% Aldrich) and diphenylether (10 mL, 99\%, Fluka) and heated to 60$^\circ$C for 90 min under oil pump vacuum on which a transparent solution was obtained. (To ensure that acetic acid is absent in the reaction mixture at this point, it is essential to apply a reduced atmosphere of $<$0.1 mbar!) The flask was flooded with nitrogen and set to ambient pressure. It was then heated to 170$^\circ$C upon which trioctylphosphine (2.50 mL, 90\%, Aldrich) was added. Acetic acid (20 $\mu$L, 99\% Aldrich) was added, immediately followed by the quick injection of (I) (2.50 mL, 1.25 mmol) upon which it was instantaneously cooled to 150$^\circ$C. (Note: For the synthesis of cubic PbTe nanoparticles, the addition of acetic acid has to be skipped.) The reaction was terminated by cooling to room temperature after 3.5 h.

\textbf{Purification of PbTe Nanoparticles for Characterization (step 3).} A fraction of the brown-yellow solution obtained under step 2 was mixed with ethanol (25 vol\%, analytical grade, Fluka) and centrifuged at 4500 rpm for 5 min. The light yellow supernatant was removed under nitrogen, and the almost black precipitate
was suspended in a few drops of chloroform (analytical grade, Fluka) on which the washing cycle was repeated two more times. The purified PbTe nanoparticles should be stored in the absence of oxygen to prevent aging.

\textbf{Ligand Exchange (Oleic Acid $\rightarrow$ Oleylamine) (step 4).} The purified (II) was precipitated again by the addition of excess ethanol, the suspension was centrifuged, and the supernatant was removed. The black precipitate was suspended in oleylamine (2 mL, 70\% Aldrich) and allowed to rest for 60 min. The mixture
was centrifuged, the supernatant was removed, and the black precipitate was dissolved in chloroform. This procedure was repeated once with a resting time in fresh oleylamine of 10 min.

\textbf{Preparation of a 1.000 M Solution of Selenium in TOP (Se$@$TOP) (step 5, product (V)).} In a glovebox, selenium (1.579 g, 20.00 mmol, 99.999\%, Chempur) was suspended in distilled TOP (20.0 mL, 90\%, Merck) under stirring. It was heated to 200$^\circ$C until a completely transparent, colorless solution was obtained. The
solution was stored in the glovebox.

\textbf{Synthesis of PbTe-PbSe Nanostars (step 6, product (VI)).} Lead acetate trihydrate (0.785 g, 2.07 mmol, 99\% Aldrich) was mixed with oleic acid (1.50 mL, 70\% Aldrich) and diphenylether (10 mL, 99\%, Fluka). The solution prepared under (step 4) of oleylaminestabilized PbTe nanoparticles as obtained under step 1 was added, and it was heated to 60$^\circ$C for 90 min under oil pump to remove chloroform, acetic acid and H$_2$O. (To ensure that acetic acid is absent in the reaction mixture at this point, it is essential to apply a reduced atmosphere of $<$0.1 mbar!) The flask was flooded with nitrogen, set to ambient pressure and it was heated to 170$^\circ$C on which trioctylphosphine (3.0 mL, 90\%, Aldrich) was added, followed by acetic acid (20 $\mu$L, 99\% Aldrich). With a syringe pump, (V) (3.0 mL, 3.0mmol) was slowly added to the dark-brown solution within 30 min (rate: 6.0 mL/h). After complete injection, the reaction temperature was lowered to 150$^\circ$C. The reaction was terminated by cooling to room temperature after 18 h. The PbTe-PbSe nanostars were purified as described under step 3.

\textbf{Ligand Removal from PbTe- and PbTe-PbSe Nanostructures (step 7, product (VII)).} The purified (II) or (VI) was precipitated with ethanol, the supernatant was removed after centrifugation, and the black precipitate was dried under vacuum. In a glovebox under nitrogen, the black solid was mixed with tetradecylphosphonic
acid (20 mg, 98\% Merck) as well as chloroform (2 mL, analytical grade, Aldrich) and allowed to stir overnight, where upon a black suspension was formed. The supernatant was removed, and it was washed three times with chloroform. The precipitate was suspended in a solution of NH$_3$ in methanol (2 mL, 7 mol L-1, Aldrich). After the mixture was stirred overnight, the supernatant was removed after centrifugation (4500 rpm, 5 min) and it was washed two times with fresh NH$_3$. Then, the precipitate was suspended in methanol (1mL, analytical grade, Aldrich) and hexane (1 mL, analytical grade, Aldrich), followed by the addition of oxygen-free acetic acid (20 droplets). The hexane phase (the upper phase) was removed, and fresh hexane was added. This procedure was repeated until the hexane phase remained clear and colorless. All solvents were removed and the precipitate was washed three times with methanol. On drying under vacuum overnight, a fine black powder was obtained. Typically, the starting amounts specified under step 2 yield approximately 150 mg of PbTe nanoparticles and 380 mg of PbTe-PbSe nanostars for step 6, respectively.

\textbf{Compaction of PbTe-Nanoparticles and PbTe-PbSe Nanostars to Pellets by Spark Plasma Sintering (step 8).} Typically, 119 mg of (VII) kept under argon was loaded into a WC/Co die of 8.0 mm x 1.5 mm in area. The powder was pressed to a solid pellet of equal dimensions and approximately 1.5 mm in height by spark plasma sintering in a SPS-515 ET/M apparatus (Dr. Sinterlab). For thermal conductivity measurements, 242 mg of (VII) was loaded into a disk-shape die of 6 mm in diameter to obtain a tablet of PbTe or PbTe-PbSe nanoparticles with 1.3 mm in height. On applying 340 MPa (for rectangular bars) or 530 MPa (for disks) pressure, the die containing the nanopowder was heated from 20 to 100$^\circ$C in 10.0 min with no hold time by applying a DC current between 0 and 165 A and immediately allowed to cool down to room temperature. The obtained PbTe or PbTe-PbSe nanoparticle pellets were mechanically robust and silver-metallic in appearance..

(HR-)TEM imaging was performed with a JEOL JEM 2200 FS (UHR) with CESCOR and CETCOR corrector at an acceleration voltage of 200 kV or a JEM-Jeol-1011 microscope at 100 kV with a CCD camera. SEM images were obtained on a LEO1550 scanning electron microscope with a spatial resolution of $\sim$1 nm. XRPDs were recorded using a Philipps Pert-diffractometer with Bragg-Brentano geometry on applying copper-K$\alpha$ radiation ($\lambda$ = 154.178 pm, U = 45 kV; I = 40 mA).

For measurements of the thermopower and resistivity a ZEM-3 apparatus (ULVAC-RIKO) was applied under a low-pressure helium atmosphere. The thermopower was determined by a static dc method where the resistivity was simultaneously measured by a four-terminal setup.

The specific heat was measured by a relaxation technique in a Physical Property Measurement System by Quantum Design. A heat pulse of 2\% of the bath temperature has been applied and repeated 3 times at each temperature.

Thermal diffusivity measurements were recorded with a Netzsch LFA-457 Microflash with a Pyroceram standard for calibration.


\end{document}